%% file: main.tex
\documentclass[sigconf,nonacm]{acmart}

\renewcommand\footnotetextcopyrightpermission[1]{} 
\AtBeginDocument{%
  }

\newcommand{\ignore}[1]{}

\input{definition}

\begin{document}

\title{ChainFuzzer: Greybox Fuzzing for Workflow-Level Multi-Tool Vulnerabilities in LLM Agents}

\author{Jiangrong Wu}
\email{wujr28@mail2.sysu.edu.cn}
\affiliation{%
  \institution{Sun Yat-sen University}
  \city{Zhuhai}
  \state{Guangdong}
  \country{China}
}

\author{Zitong Yao}
\email{yaozt@mail2.sysu.edu.cn}
\affiliation{%
  \institution{Sun Yat-sen University}
  \city{Zhuhai}
  \state{Guangdong}
  \country{China}
}

\author{Yuhong Nan}
\authornote{Corresponding author.}
\affiliation{%
  \institution{Sun Yat-sen University}
  \city{Zhuhai}
  \state{Guangdong}
  \country{China}
}
\email{nanyh@mail.sysu.edu.cn}

\author{Zibin Zheng}
\affiliation{%
  \institution{Sun Yat-sen University}
  \city{Zhuhai}
  \state{Guangdong}
  \country{China}
}
\email{zhzibin@mail.sysu.edu.cn}

\renewcommand{\shortauthors}{Wu et al.}

\begin{abstract}
Tool-augmented LLM agents increasingly rely on multi-step, multi-tool to complete real tasks. This design expands the attack surface, because data produced by one tool can be persisted and later reused as input to another tool, enabling exploitable source-to-sink dataflows that only emerge through tool composition. We study this risk as multi-tool vulnerabilities in LLM agents, and show that existing discovery efforts that focus on single-tool or single-hop testing can miss these long-horizon behaviors and provide limited debugging value.

We present ChainFuzzer, a greybox framework for discovering and reproducing multi-tool vulnerabilities with auditable evidence. ChainFuzzer (i) identifies high-impact operations with strict source-to-sink dataflow evidence and extracts plausible upstream candidate tool chains based on cross-tool dependencies, (ii) uses Trace-guided Prompt Solving (TPS) to synthesize stable prompts that reliably drive the agent to execute target chains, and (iii) performs guardrail-aware fuzzing to reproduce vulnerabilities under LLM guardrails via payload mutation and sink-specific oracles.

We evaluate ChainFuzzer on 20 popular open-source LLM agent apps (998 tools). ChainFuzzer extracts 2,388 candidate tool chains and synthesizes 2,213 stable prompts, and confirms 365 unique, reproducible vulnerabilities across 19/20 apps, where 302 require multi-tool execution. In component evaluation, tool-chain extraction achieves 96.49\% edge precision and 91.50\% strict chain precision; TPS increases chain reachability from 27.05\% to 95.45\%; and guardrail-aware fuzzing increases the payload-level trigger rate from 18.20\% to 88.60\%. Overall, ChainFuzzer achieves an efficiency of 3.02 vulnerabilities per 1M tokens, providing a practical foundation for testing and hardening real-world multi-tool agent systems.
\end{abstract}



\maketitle

\input{main-body}

\bibliographystyle{ACM-Reference-Format}
\bibliography{main}

\typeout{get arXiv to do 4 passes: Label(s) may have changed. Rerun}
\end{document}

%% file: definition.tex
\newcommand{\para}[1]{\vspace{1pt}\noindent\textbf{#1.~}}
\newcommand{\system}{\textit{\sloppy{ChainFuzzer}\@}}
\usepackage{cleveref}
\usepackage{xcolor}
\usepackage{colortbl}
\usepackage{multirow}
\usepackage{fancyvrb}  
\usepackage{mdframed}  
\usepackage{xcolor}
\usepackage{algorithm}
\usepackage{algpseudocode}
\usepackage{amsmath} 
\usepackage{setspace}
\usepackage{tcolorbox}  
\usepackage{xcolor}     
\usepackage{lipsum}     
\usepackage{graphicx}
\usepackage{makecell}
\usepackage{tabularx} 
\newtcolorbox{graybox}{
    colback=gray!20,    
    colframe=white,     
    boxrule=0pt,        
    width=\textwidth,   
    left=1em,           
    right=1em,          
    top=0.5em,          
    bottom=0.5em,       
    breakable           
}

\usepackage{enumitem}
\usepackage{pifont}
\usepackage{multirow}
\usepackage{siunitx}

%% file: main-body.tex
\section{Introduction}
\label{sec:intro}

Large language models (LLMs) are increasingly used as \textit{tool-augmented agents} that can autonomously execute tasks rather than only answer questions~\cite{agent-argumented-paper,agent-argumented-survey}. By integrating with external tool ecosystems (e.g., web search, file systems, databases, and code runners), an agent can interpret a user request, plan sub-tasks, and coordinate data flow across multiple components in a closed loop. This capability enables complex automation, such as searching for a repository, downloading a file, extracting key information, and applying a patch by chaining tool invocations (e.g., \texttt{web\_search} $\rightarrow$ \texttt{download} $\rightarrow$ \texttt{write\_file} $\rightarrow$ \texttt{run}). However, the same multi-stage dataflow also expands the attack surface: information produced by one tool can become the input of another tool in later steps, and the agent may unintentionally propagate attacker-influenced content into high-impact operations.

In this paper, we study this security risk as \textbf{multi-tool vulnerabilities} in LLM agents, which arise when tool composition creates an exploitable source-to-sink dataflow across multiple steps. Currently, LLM agents increasingly rely on \emph{multi-tool workflows} to complete real tasks, which makes vulnerability logic more complex and often hidden behind intermediate steps and artifacts. In practice, an unsafe behavior may only emerge after several tool calls and state transitions (e.g., a value fetched from the web is written to a file and later reused as an input to an execution tool), so assessing each tool in isolation does not capture the security boundary of the agent as a whole.

However, existing vulnerability discovery efforts~\cite{agentfuzz, LLMSmith,taintp2x,single-hop-jailbreaking} for LLM agents focus on \emph{single-tool} or \emph{single-hop} testing, where the goal is to directly trigger one unsafe tool call with an obviously malicious input. This is insufficient in real ecosystems for two reasons. First, it leads to substantial false negatives: multi-step chains can bypass this testing model even when every step looks benign in isolation, and the sink effect is only reachable through composition. Second, single-tool testing provides limited debugging value for developers: it cannot explain how attacker-influenced content traverses across tools and carriers, and therefore cannot guide mitigation at the right boundary (e.g., the first untrusted ingress point or the downstream sink parameter). As a result, there is a need of a workflow-level testing approach that can model, execute, and validate multi-tool chains, and provide auditable evidence to improve the security of deployed agent systems.

\para{Our Work}
In this paper, we propose \system{}, a greybox automated and generic framework for discovering \emph{multi-tool} vulnerabilities in LLM agents. Given an agent app (its tool set, tool schemas/descriptions, and optionally tool source code), \system{} aims to produce \emph{auditable and reproducible} vulnerability that can be verified by developers.

To achieve this goal, we must overcome three practical challenges: \textit{(1) Finding risky tool chains; (2) Driving the agent to execute a target chain; (3) Validating sink effects under guardrails}. 
First, risky behaviors are embedded in a large space of possible tool combinations, and naive exploration wastes budget on semantically invalid or irrelevant sequences. We need to efficiently recover realistic upstream chains that can deliver attacker-influenced content to high-impact operation.
Second, even when a candidate chain is known, an LLM agent is not a deterministic executor: small prompt changes can lead to different plans, tool choices, and argument bindings, or fail due to missing preconditions, making it difficult to reliably reproduce a specific workflow. 
Third, validation must work under model guardrails that often refuse or weaken obviously malicious payloads; this can hide compositional weaknesses of the agent unless we can test in a guardrail-aware way and still reproduce the vulnerability.

\system{} addresses these challenges in three steps. First, it identifies high-impact \texttt{sink\_tools} with strict source-to-sink dataflow evidence and extracts plausible upstream \texttt{candidate\_chains} based on cross-tool dependencies. Second, it uses Trace-guided Prompt Solving (TPS) to synthesize a stable \texttt{valid\_prompt} for each chain by comparing runtime traces against the target chain and iteratively repairing the prompt until the agent can execute the chain reliably. Third, it performs guardrail-aware fuzzing to reproduce vulnerabilities under LLM guardrail by mutating the payloads and validating exploitation with sink-specific oracles.

We evaluate \system{} on a dataset of \num{20} popular LLM agents, which together expose \num{998} tools. From these tool ecosystems, \system{} extracts \num{2,388} candidate tool chains and synthesizes \num{2,213} stable prompts, enabling systematic end-to-end testing of long-horizon multi-tool chains. Our findings show that multi-tool vulnerabilities are common in real agent deployments: \system{} confirms \num{365} unique, reproducible vulnerabilities across \num{19}/\num{20} apps, and \num{302} of them are multi-tool vulnerabilities, indicating that most real risks require cross-tool composition rather than a single unsafe tool invocation.
From a technical perspective, our component results validate that guided reachability and guardrail-aware reproduction are necessary for reliable discovery. On a manually validated sample of extracted tool-chain graphs, \system{} achieves high precision at both the edge level (\num{96.49}\%) and the strict chain level (\num{91.50}\%), indicating that the extracted chains largely correspond to real, semantically meaningful dependencies. TPS substantially improves executability: reachability increases from \num{27.05}\% to \num{95.45}\%. Finally, guardrail-aware fuzzing is essential for reproduction under safety filters: the overall payload-level trigger rate increases from \num{18.20}\% to \num{88.60}\% with mutation.

\system{} provides practical security benefits for agent developers and platform operators. For developers, \system{} produces actionable and reproducible vulnerabilities/PoCs that improve the security boundary and robustness of the LLM agent. To faciliate future research, we have released the corresponding artifact: \url{xxx}.

\para{Contribution}
In summary, this paper makes the following contributions:

\vspace{-6px}
\begin{itemize}[leftmargin=*]
    \item We identify and characterize \emph{multi-tool vulnerabilities} in LLM agents, where attacker-influenced content propagates across tool boundaries and reaches high-impact operations.
    
    \item We develop \textbf{\system{}}, a greybox testing framework that combines sink-related tool-chain extraction, Trace-guided Prompt Solving (TPS), and guardrail-aware fuzzing to produce auditable and reproducible PoCs (prompt, tool-call trace, and evidence).
    
    \item We conduct an extensive evaluation on 20 popular agents and show that multi-tool vulnerabilities are common and practically discoverable: \system{} finds 365 unique vulnerabilities across 19/20 apps, where 82.74\% require multi-tool execution, and achieves an efficiency of 3.02 vulnerabilities per 1M tokens.
\end{itemize}

\section{Background and Problem Statement}
\label{sec:background}

\subsection{LLM Agent Background}
\label{sec:arch}

A typical LLM-based agent integrates a backbone LLM with a set of external tools, such as file system operations, database queries, web APIs, and code execution environments. Given a user request, the agent executes an iterative loop: the LLM interprets the request, decides the next action, invokes a tool with concrete arguments, and then uses the tool output to continue the task. The key difference from a traditional scripted workflow is that tool selection and argument construction are decided at runtime by the LLM, so the exact sequence of tool calls can vary across runs, even under similar inputs.

\para{Execution LLM}
In practice, the LLM plays the role of a controller that coordinates multi-step execution. Concretely, it first decomposes a high-level request into tool-level subgoals, then the LLM selects tools based on their descriptions and current context, and finally binds tool arguments using intermediate results produced during execution. The whole process can be represented as an \emph{execution trace}, i.e., an ordered sequence of tool invocations with their arguments and outputs. This trace view is important because many real tasks require more than one tool call, and small deviations in one step (e.g., choosing a different tool or using a different field from a prior output) can change what happens in later steps.

\para{Tool interfaces}
From a software engineering perspective, each tool can be modeled by its interfaces and its interaction with the external environment. A tool typically exposes (1) an input schema (Function signature or JSON schema) that defines required fields and formats; (2) an output object that may contain structured fields rather than plain text; and (3) possible side effects on the environment. Side effects are common in agent deployments because tools are often used to perform real actions, not only to compute values. For example, a file tool may create or overwrite a file, a database tool may update a record, and a network tool may send an HTTP request to a remote endpoint. These interface and effect properties are the basis for reasoning about how tool calls can be composed into larger workflows.


\para{Multi-tool dependencies}
Multi-tool workflows are characterized by dependencies where intermediate artifacts produced by one tool are consumed by later tools. There are two common dependency patterns. First, \emph{direct output-to-input dependencies} occur when a later tool directly takes a field from a previous tool's output as its argument, possibly after lightweight transformation by the LLM. For example, a web search tool may return a URL field, and a download tool may use that URL as its input argument (e.g., \texttt{search} returns \texttt{url}, then \texttt{download(url)}). Second, \emph{indirect dependencies} occur when an intermediate artifact is persisted in a shared carrier and later retrieved or referenced by another tool. Common carriers include files, database records, retrieval indexes, and caches. For example, the agent may download content, write it to a local file, and later invoke a code execution tool that reads and runs that file (e.g., \texttt{download} returns \texttt{content}, \texttt{write\_file(path, content)} persists it, then \texttt{run(path)} consumes \texttt{path}). These multi-tool dependencies are essential for functionality and expand the reachable execution space of an agent; however, they also create long, cross-step execution paths where untrusted or user-influenced artifacts can be transformed and reused before reaching downstream actions, which makes workflow-level security failures possible.

\subsection{Problem Statement}
\label{sec:problem}

\begin{figure}[htbp]
    \centering
    \includegraphics[width=0.45\textwidth]{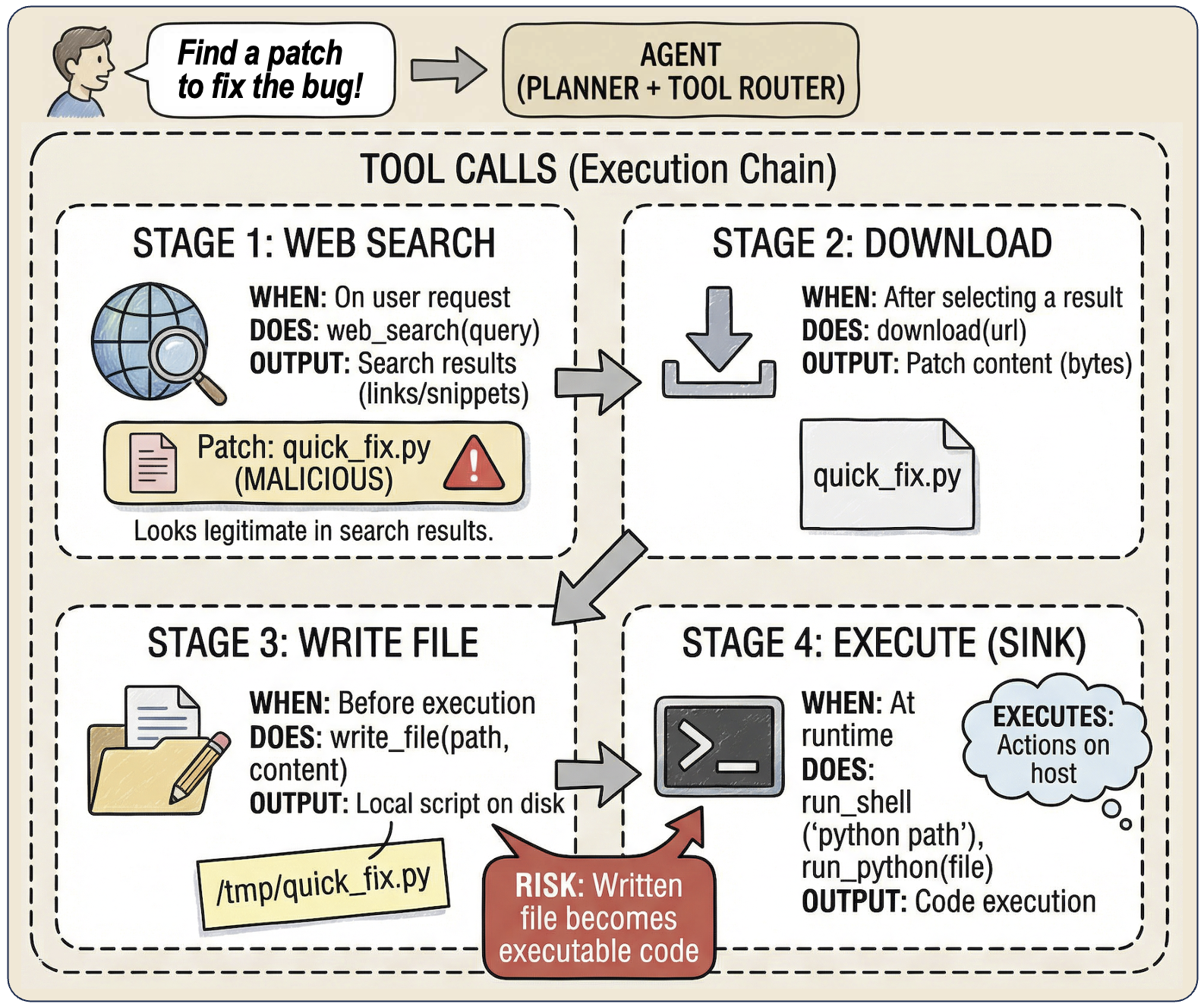}
    \caption{Motivating example of a multi-tool workflow that turns untrusted web content into executable code via \texttt{web\_search} $\rightarrow$ \texttt{download} $\rightarrow$ \texttt{write\_file} $\rightarrow$ \texttt{execute}.}
    \label{fig:motivating_example}
\end{figure}

\para{Motivating example}
Figure~\ref{fig:motivating_example} shows a common multi-tool workflow in modern agents. A user asks the agent to ``find a patch to fix a bug.'' The agent first uses a web search tool to obtain candidate results, then downloads one result, writes the downloaded content to a local file, and finally executes the file using a code execution tool. Each step is individually reasonable for the stated task: searching for a patch, downloading it, saving it as a script, and running it to apply the fix. The security risk comes from the \emph{composition}: if the downloaded content is untrusted, the workflow can promote that content into executable code and trigger unintended command or code execution (e.g., the search result points to a script that contains a hidden payload, and the agent runs it as a ``quick fix'').

We study \emph{workflow-level vulnerabilities} in tool-augmented LLM agents. In these cases, the unsafe behavior is not caused by a single tool call in isolation, but by a multi-step tool chain where intermediate artifacts (e.g., URLs, code snippets, downloaded files) are created and reused across tools. Beyond untrusted web content, multi-tool ecosystems enable many other vulnerability patterns that only emerge at the workflow level. For example, a malicious local user can steer a remote, cloud-hosted agent (e.g., Chatgpt, Gemini, Doubao) service to combine tools in an unsafe way, where each individual step looks benign but the overall chain becomes dangerous. These examples show that multi-tool ecosystems increase agent capability, but they also expand the attack surface through longer dependency chains and artifact reuse.

However, the existing agent vulnerability discovery techniques~\cite{agentfuzz,LLMSmith} remain centered on \emph{single-tool} or \emph{single-hop} testing, where one prompt is evaluated against one tool call invocation, even though real-world agent applications most commonly rely on \emph{collaborative multi-tool workflows} to complete tasks. Such approaches under-explore multi-step tool dependencies and therefore miss risks that only materialize after cross-tool dataflow and state changes, leaving workflow-level vulnerabilities largely untested in practice.

\para{Research goal}
In this paper, our goal is to help developers uncover deep vulnerabilities in LLM agents with multi-tool operation. We build an automated, greybox testing framework that \textit{(i) models multi-tool workflows} so that tool dependencies and intermediate artifact propagation can be analyzed at the workflow level, and \textit{(ii) reproduces workflow-level vulnerabilities} by generating test inputs that drive the agent to execute specific tool chains with malicious payload. Finally, we generate minimal, auditable PoCs with tool-call traces and effect evidence. By achieving this framework, we aim to improve the security and robustness of real-world multi-tool agent systems. 

\para{Scope and assumptions}
We focus on vulnerabilities that arise from tool compositions in deployed agent settings, where tools can ingest untrusted content from external sources or shared carriers (e.g., web pages, downloaded files, database records), and where a malicious user can also directly steer a remote agent through natural-language interactions to execute unsafe multi-tool workflows. Our framework uses gerybox testing to simulate such untrusted inputs and to make executions reproducible; we do not assume an attacker can modify the agent code or tool implementations in deployment. We do not consider threats that require access to model weights, developer secrets, or system privileges beyond the permissions already granted to the agent tools.


\section{System Overview}
\label{sec:overview}

\subsection{Challenges and Key Ideas}
\label{sec:challenges}

\para{Challenges}
To achieve workflow-level vulnerability discovery in multi-tool LLM agents faces three practical challenges:

\begin{itemize}[leftmargin=*]
    \item \textbf{Challenge-1: Finding risky tool chains.}
    In multi-tool agents, risky behaviors are often embedded in a large space of possible tool combinations. With a large and heterogeneous tool set, it is difficult to efficiently narrow down to \emph{realistic} upstream chains that can lead to high-impact actions, rather than spending budget on semantically invalid or irrelevant sequences. This requires capturing accurate data dependence between tools and find the potential vulnerable tool chains. 

    \item \textbf{Challenge-2: Driving the agent to execute the tool chain.}
    Even when a candidate chain is known, reliably reproducing it is difficult because an LLM agent is not a deterministic executor. Small changes in prompts can lead to different plans, different tool choices, different argument bindings, or runtime failures, which makes it hard to force the agent to follow a specific sequence. Common issues include missing preconditions, path/name mismatches for intermediate artifacts, and format or parameter mismatches across steps. This situation requires generating a valid prompt that can drive the agent to execute the specific tool chain during the vulnerability verification. 

    \item \textbf{Challenge-3: Reproducing vulnerability under guardrails.}
    During reproduction, intrinsic LLM guardrails can refuse or sanitize explicit malicious payloads, reducing the effectiveness of straightforward testing and potentially hiding agent tool-level design weaknesses. While this can provide partial protection in deployed agents, it may also encourage over-reliance on the LLM as a primary defense. For vulnerability validation, we need to reproduce unsafe tool effects in a way that bypass the LLM native guardrails, so that tool-level design weaknesses can be exposed.
\end{itemize}



\begin{figure*}[htbp]
    \centering
    \includegraphics[width=0.95\textwidth]{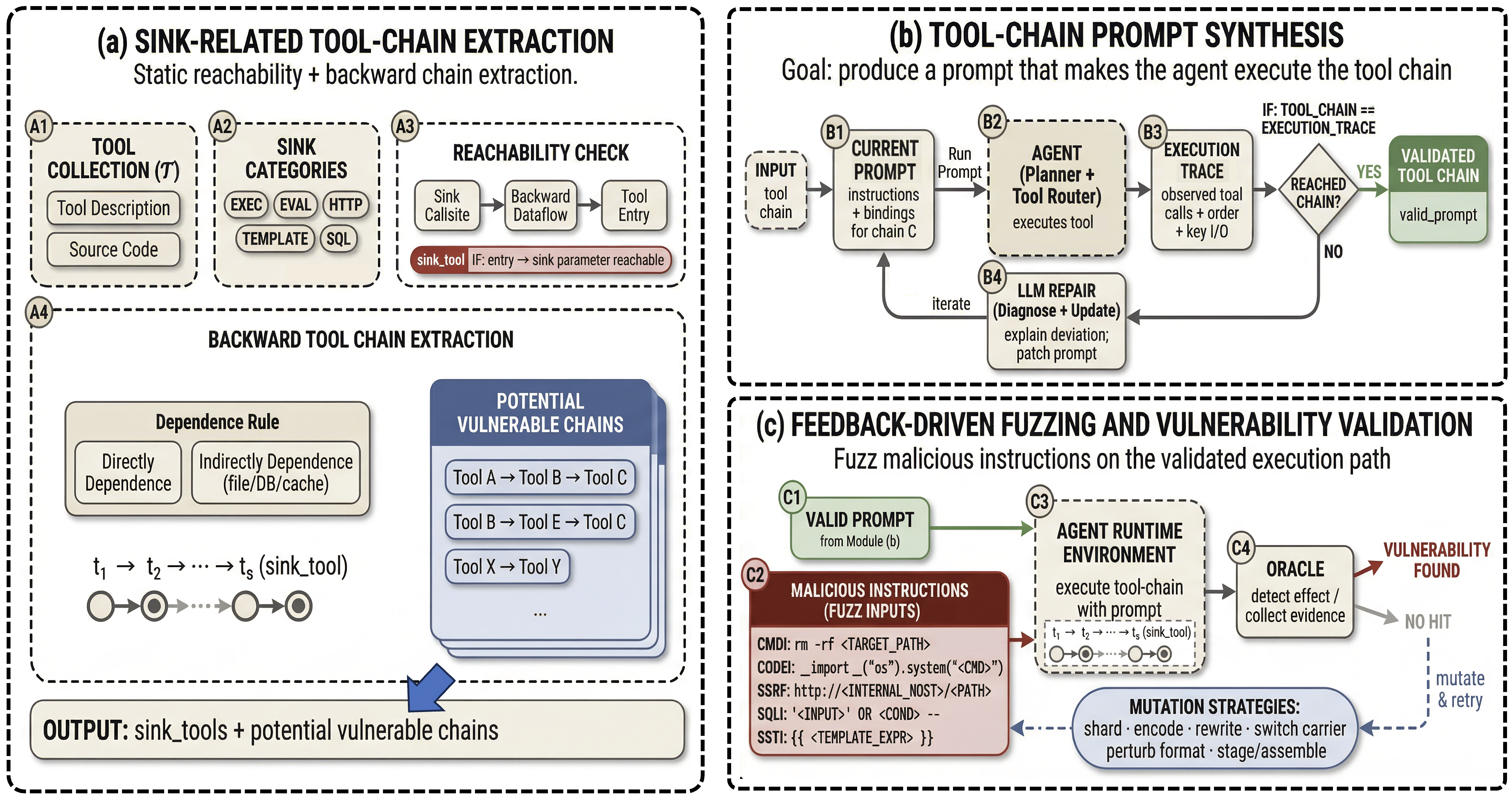}
    \caption{Workflow of \system{}.}
    \label{fig:workflow}
\end{figure*}

\subsection{\system{} Workflow}
\label{sec:system_overview}

\system{} is a greybox testing framework for discovering and reproducing workflow-level vulnerabilities in multi-tool LLM agents. Given an agent's tool set (tool interfaces, description and source code), \system{} outputs reproducible vulnerability proof-of-concepts (PoCs) with the triggering prompt, malicious payload, and tool-call traces. \system{} first identifies candidate risky workflows from tool dependencies, then drives the agent to execute those workflows reliably, and finally validates high-impact unsafe effects under realistic guardrail settings. As shown in Figure~\ref{fig:workflow}, \system{} consists of three modules: \textit{Sink-related tool-chain extraction}, \textit{Tool-chain prompt generation}, and \textit{Feedback-driven fuzzing \& vulnerability validation}.

\para{Sink-related tool-chain extraction}
To overcome challenge-1, \system{} first reduces the search space by focusing on high-impact endpoints. It scans tool code to identify \texttt{sink\_tool}s using a strict source-to-sink reachability rule: a tool is marked as a sink only if tool inputs can influence the arguments of a high-risk API call (e.g., exec, eval). Starting from each \texttt{sink\_tool}, \system{} then performs backward chain extraction to identify upstream tools that can plausibly feed data or artifacts into the \texttt{sink\_tool}. The extraction captures both direct output-to-input dependencies and indirect dependencies through shared carriers such as files or database records. The output of this module is a set of potential vulnerable chains, each annotated with a dependency rationale and the key artifact/field that propagates to the sink argument.

\para{Tool-chain prompt generation}
For each candidate chain, \system{} must reliably drive the agent to execute the intended tool sequence before any vulnerability can be validated. To do so, \system{} uses \emph{Trace-guided Prompt Solving (TPS)} to synthesize an executable prompt conditioned on the chain. It starts from a seed prompt derived from the chain and runs the agent to collect a tool-call trace. If the execution diverges, \system{} analyzes the trace to determine where and why the chain was not followed (e.g., missing preconditions, incorrect bindings, or runtime failures), and uses this constraint to revise the prompt. This trace-guided repair iterates until the chain becomes reachable and stable. The output is a \texttt{valid\_prompt} that consistently reproduce the target chain to address challenge-2.

\para{Feedback-driven fuzzing \& vulnerability validation}
Once a chain is executable, \system{} validates whether it can trigger a security-relevant effect at the sink under realistic guardrail settings. This module generates attack payloads based on the sink type (e.g., command execution, SSRF, SQL, template execution) and selects a payload injection point based on the dependency annotation from Module~(a), so that the payload is placed on the tool chain path that reaches the sink argument. \system{} supports two injection channels: user-driven injection, where the payload is introduced via natural-language inputs, and environment-driven injection, where the payload is introduced via untrusted external content or shared carriers that the agent consumes. When direct payloads are blocked or degraded by intrinsic LLM guardrails (challenge-3), \system{} applies feedback-driven mutation to bypass the guardrails through adjusting the payload form while keeping the overall workflow execution. Finally, \system{} applies sink-specific oracles based on observable tool effects and trace evidence, producing a minimal PoC trace for reporting.

\section{Design Details of \system{}}
\label{sec:design}

\subsection{Sink-Related Tool-Chain Extraction}
\label{sec:tdg}

This module extracts candidate \emph{risk chains} by working backward from high-impact tool capabilities. The input is an agent tool set $\mathcal{T}$, where each tool provides an interface (name, parameters, return fields), description, and source code. The output consists of (1) a set of \texttt{sink\_tools} whose implementations contain high-impact sink APIs with \emph{input-reachable} sink arguments, and (2) a set of \texttt{candidate\_chains} that end at each \texttt{sink\_tool} and satisfy multi-tool dependency constraints. Each extracted chain is annotated with its dependency type (direct or indirect) and an optional \emph{key dependency site} that indicates which upstream artifact is most likely to influence a sink argument.

\input{sink_type}

\subsubsection{Sink tool identification}
We identify \texttt{sink\_tools} by combining a sink API catalog with strict intra-tool source-to-sink dataflow analysis. We first match each tool's source code against the sink API catalog in Table~\ref{tab:sink_api_table_compact_lines} to find sink callsites and their corresponding sink types (CMDi, CODEi, SSRF, SSTI, SQLi). We then apply a strict reachability rule: a tool is labeled as a \texttt{sink\_tool} only if at least one sink callsite receives an argument that can be influenced by the tool's entry inputs. In practice, we implement this check using backward data flow over the tool code, starting from the sink argument expression and tracing through common code structures (e.g., assignment chains, intermediate variables and so on). This rule intentionally excludes tools that merely contain sink calls with fixed constants, which reduces false positives and keeps later validation focused on tools where tool inputs or environment-derived artifacts can actually reach high-impact actions.

\subsubsection{Backward chain extraction}
Starting from each identified \texttt{sink\_tool}, we recover upstream tool chains using \textit{static dependency analysis} followed by \textit{LLM-based semantic filtering}.

We construct dependency edges with two static rules:
\begin{itemize}[leftmargin=*]
    \item \textbf{Direct dependencies.} We connect two tools when an upstream tool's return field can be consumed by a downstream tool's parameter under an interface-level \emph{equivalence} or \emph{containment} relation. Equivalence covers compatible fields such as \texttt{url} or \texttt{path}; containment covers consuming a sub-field from a structured output (e.g., a returned object containing a \texttt{url} field).

    \item \textbf{Indirect dependencies.} We connect two tools when the upstream tool can persist content (e.g., via file/DB write APIs) and the downstream tool can later read or execute persisted content (e.g., via file/DB read APIs or execution APIs that load a file).
\end{itemize}

These rules intentionally over-approximate the tool-chain space and may produce interface-compatible but semantically implausible chains. We therefore use an LLM as a semantic filter: given a candidate chain, the LLM judges whether the chain forms a coherent workflow that an agent would plausibly execute to complete a realistic task, and discards chains that lack logical linkage between steps. For example, a chain that connects a tool that \emph{summarizes a document} to a tool that \emph{creates a database index} may be type-compatible (both manipulate strings) but is not a reasonable sequence for a user task. After filtering, we retain a set of sink-related potential vulnerable chains for next modules.

\subsection{Tool-Chain Prompt Generation}
\label{sec:prompt generation}

Given a candidate tool chain $C=\langle t_1,\dots,t_k\rangle$ extracted in \Cref{sec:tdg}, this module generates an executable prompt that reliably drives the agent to follow $C$. The module takes as input the chain $C$ together with tool interfaces and descriptions, and outputs a valid prompt and its corresponding trace. The key difficulty is that tool-using agents are non-deterministic planners: small prompt changes can lead to different tool choices, different argument bindings, and different runtime outcomes. In our setting, divergences between the observed trace and the target chain are most often caused by \emph{semantic-level mismatches in the prompt}, such as incorrect bindings, missing preconditions, or ambiguous step requirements. \system{} therefore uses an LLM, which provides strong semantic understanding and controllable text generation, to construct trace-derived constraints and to solve for prompt revisions.

\para{Seed prompt generation}
\system{} leverages LLM to generate an initial seed prompt $P_0$ conditioned on the target chain $C$ and tool metadata (tool descriptions and interface information). The seed prompt is a step-by-step user request that instructs the agent to invoke the tools in the order of $C$ to complete a plausible task.

\subsubsection{Constraint generation from traces}
\system{} executes the agent with the current prompt $P_i$ and records a trace $\tau_i$ via instrumentation that summarizes the tuple $\langle m, t, a, r, e \rangle$, where $m$ is the thinking of the LLM of the agent, $t$ is the invoked tool, $a$ is the argument of the tool, $r$ is the return value, and $e$ is the execution status (e.g., success, exception type, or timeout). \system{} then uses an LLM-based constraint generator to compare $\tau_i$ with the target chain $C$ and produce a compact set of semantic constraints $\Delta_i$ that explain why the execution failed to follow $C$. The constraints focus on actionable, prompt-level causes, including \textbf{\textit{(i) missing preconditions}} (e.g., a required file, credential, or record was never created), \textbf{\textit{(ii) incorrect or incomplete argument bindings}} (e.g., a later step did not use the artifact produced earlier), \textbf{\textit{(iii) format or parameter mismatches}} (e.g., a tool expects a file path but received a string of the path), \textbf{\textit{(iv) permission or confirmation gates}} (e.g., the agent requests user approval before invoking a tool), and \textbf{\textit{(v) planner detours}} (e.g., the agent chose a different tool or skipped a required step).

\subsubsection{Prompt solver}
Given $(P_i, \Delta_i)$, \system{} uses an LLM-based solver to produce a revised prompt $P_{i+1}$ that satisfies $\Delta_i$ while preserving as much of the already-correct prefix as possible. Formally, the solver searches for a feasible update within a constrained local-edit space: $P_{i+1}\in \mathrm{Edit}(P_i,\Delta_i)\ \text{s.t.}\ P_{i+1}\models \Delta_i$, where $P\models \Delta_i$ means the prompt explicitly encodes the semantic constraints extracted from the trace (e.g., required steps, bindings, and preconditions), and $\mathrm{Edit}(P_i,\Delta_i)$ denotes a restricted set of local prompt edits around the failure point (e.g., inserting a missing step, strengthening one instruction, or rewriting an ambiguous binding), rather than a full prompt rewrite.

The solver applies local edits that directly address the reported failure modes. Typical edits include adding a missing step, strengthening an instruction that the agent previously ignored, clarifying an ambiguous reference or binding, and repairing missing preconditions by revising the prompt to explicitly invoke the tool step(s) that create the required resource (e.g., if a later step expects a local file path, the prompt is revised to require calling the file-writing tool to create that file before the execution step). \system{} iterates this repair loop until either the agent reaches and executes the full chain $C$ consistently, or a fixed budget is exhausted (maximum iterations, tool calls, or runtime). For the cost consideration and stability trade-off, if the agent executes the full chain $C$ successfully for five consecutive runs, \system{} treats the prompt as stable and accepts it as a valid prompt.

\subsection{Fuzzing-Driven Vulnerability Validation}
\label{sec:mutation}

This module validates whether a reachable tool chain can be affected by the vulnerability. The input is a valid prompt produced in \Cref{sec:prompt generation}, and the malicious payload. The output is an auditable PoC: a triggering prompt, the resulting tool-call trace, and evidence that a sink-level unsafe effect occurred. The main challenge is make the malicious payload bypass the native guardrails of LLM and validate the vulnerability to expose the risk of the multi-tool level design.

\subsubsection{Injection point selection}
Given a reachable tool chain $C=\langle t_1,\dots,t_k\rangle$ and its valid prompt, \system{} determines \emph{where} to place a malicious payload so that the chain can trigger a high-impact unsafe effect at the ending \texttt{sink\_tool}. This step follows the same real-world scope described in \Cref{sec:problem}: in deployed agent settings, risk-driving inputs typically come from (i) \textbf{untrusted external content} ingested by tools (e.g., web pages, downloaded files, database records, indexed documents), or (ii) a \textbf{malicious user} who steers a remote agent through natural-language interactions to execute an unsafe multi-tool workflow. Accordingly, \system{} treats injection point selection as deciding whether the payload should be introduced through the user dialogue or through external content that a tool ingests, and mapping that choice to a concrete boundary on the chain (i.e., the earliest step where untrusted data can enter and later propagate to the sink).

More specifically, for each reachable tool chain, \system{} determines the appropriate injection position based on the type of the first tool in the chain and the nature of untrusted data ingress. If the first tool directly receives user-provided content (e.g., \texttt{write\_file}, user-controlled fields), \system{} tests \textbf{user-source injection}, where the malicious payload is embedded in the user prompt and later passed into the tool's arguments or produced artifacts. On the other hand, if the chain involves external content (e.g., \texttt{web\_search}, \texttt{download}, reading files or database records), \system{} tests \textbf{environment-source injection}. Here, the payload is either injected into the tool’s return field or embedded into external resources (e.g., files or database entries) that the tool later reads. \system{} does not prioritize one position over the other; if both injection positions are plausible, both are tested to ensure comprehensive coverage.

\label{subsubsec:Guardrail-aware fuzzing}
\subsubsection{Guardrail-aware fuzzing}
In this step, \system{} executes the agent under instrumentation and injects the malicious payload at the planned injection point. For each reachable chain, \system{} first instantiates the malicious payload according to the sink type of the ending \texttt{sink\_tool}. Concretely, CMDi payloads are command strings that can be executed by a shell (e.g., \texttt{rm -rf <TARGET\_PATH>}); CODEi payloads are executable code snippets (e.g., Python code that invokes \texttt{os.system(...)}); SSRF payloads are URLs that target internal hosts or sensitive endpoints (e.g., \texttt{http://<INTERNAL\_HOST>/<PATH>}); SQLi payloads are query fragments that can alter query logic (e.g., tautology-style conditions); and SSTI payloads are template expressions that trigger evaluation in the template engine (e.g., \texttt{\{\{ <TEMPLATE\_EXPR> \}\}}). The exact payload string is parameterized to match the expected sink argument format (command, code, URL, query, or template).

Next, \system{} executes the agent under instrumentation and injects the malicious payload at the planned injection point. \system{} applies oracles below to decide whether the vulnerability is reproduced:

\begin{itemize}[leftmargin=*]
    \item \textbf{CMDi oracle.} The command-execution sink API is reached with payload-influenced arguments.

    \item \textbf{CODEi oracle.} The code-execution sink API is reached with payload-influenced code and the CODE probe effect is observed.

    \item \textbf{SSRF oracle.} The network-request sink API is reached with a payload-influenced destination and the request is resolved to an SSRF probe target class (e.g., internal/loopback/link-local) or a designated internal endpoint.

    \item \textbf{SQLi oracle.} The database-execution API is reached with a payload-influenced query and the SQLi probe effect is observed (e.g., unexpected acceptance/row expansion/error signature).

    \item \textbf{SSTI oracle.} The template-rendering API is reached with a payload-influenced template and the SSTI probe effect is observed in the rendered output.
\end{itemize}

If malicious payload is block by the guardrail of the LLM, \system{} mutates the payload while keeping the prompt unchanged. The mutation space uses several transformations, including (i) \textbf{shard} the payload into multiple benign-looking fragments that are reassembled later in the chain, (ii) \textbf{encode} the payload (e.g., base64 or escaping) and rely on downstream interpretation to recover it, and (iii) \textbf{perturb format} while preserving tool-required syntax (e.g., quotation, separators, and argument layout). \system{} then retry to run the agent with the mutated payload and repeats the mutation until either the vulnerability is reproduced or a fixed fuzzing budget is reached.

\section{Evaluation}
\label{sec:eval}

\input{Agent_app_experiment_object}

\input{Overall_result}

\para{Dataset}
Our dataset consists of \textbf{20 open-source, multi-tool LLM agent applications (\Cref{tab:agent_github_stars})}. We manually curated this set by searching popular GitHub repositories with agent-related keywords (e.g., ``LLM agent'', ``AI agent'') and reviewing the top results ranked by GitHub stars. We then applied two inclusion criteria: (i) each application must have at least 2{,}000 GitHub stars to ensure it is widely used and maintained, and (ii) the agent must support multi-tool workflows (i.e., it can invoke multiple distinct tools to solve different tasks). This process yields a representative set of widely adopted multi-tool agents spanning different frameworks and deployments.


\para{Environment setting}
For all evaluated agent apps, we use GPT-5.1~\cite{openai_gpt51_models} as the target LLM embedded in the agent runtime, which is one of the flagship model for agentic tasks. the LLM-related part of \system{} is driven by GPT-4o~\cite{openai2024gpt4ocard}. All experiments are conducted on a single local machine equipped with an Apple M4 Max CPU and 36\,GB RAM (Mac Pro).

\para{Research Questions} We evaluate \system{} by answering the following three research questions:
\begin{itemize}[leftmargin=*]
  \item \textbf{RQ1: Overall result and effectiveness.} How many multi-tool vulnerabilities does \system{} uncover, and what is the overall cost?
  \item \textbf{RQ2: Effectiveness of individual components.} How effective are the key modules in enabling vulnerability discovery and reproduction?
  \item \textbf{RQ3: Ablation study.} How much does each module contribute to end-to-end vulnerability yield and coverage?
\end{itemize}





\subsection{RQ1: Overall Result and Overhead}
\label{sec:rq1}

\para{Multi-tool vulnerability discovery}
\Cref{tab:overall_result} summarizes the end-to-end results of \system{} on our dataset. Across 20 agents and \num{998} tools, \system{} identifies \num{199} sink tools and constructs \num{2,388} candidate tool chains. Then \system{} synthesizes \num{2,213} prompts that can execute their target chains successfully for 5/5 repeated runs, and ultimately confirm \num{365} unique vulnerabilities affecting \num{19}/\num{20} apps.
A key observation is that workflow-level, multi-tool vulnerabilities dominate the discovered issues: \num{302}/\num{365} (\num{82.74}\%) vulnerabilities require multi-tool execution, while only \num{63} are single-tool issues. In other words, multi-tool vulnerabilities are \num{4.79}$\times$ more common than single-tool vulnerabilities, suggesting that a single-tool-only testing method~\cite{agentfuzz, LLMSmith} would miss the majority of security-relevant behaviors in deployed agents.

\Cref{tab:vuln_breakdown_source_type} further shows that vulnerabilities arise from both user-driven interactions and untrusted environment data.
Overall, \num{225} (\num{61.64}\%) vulnerabilities are triggered via user-source injection (i.e., malicious user), while \num{140} (\num{38.36}\%) are triggered via env-source injection (e.g., tool-returned web content or retrieved artifacts).
The distribution also varies by sink type: SSRF is more frequently induced by env-source inputs (\num{53} vs.\ \num{44}), which is consistent with network destinations being derived from external content (e.g., URLs embedded in retrieved pages), whereas CMDi/CODEi are more often user-source (\num{82} vs.\ \num{39} and \num{53} vs.\ \num{26}), reflecting that interactive instructions more directly steer execution parameters.

\input{Overall_result_sink_type}

\para{Cost and overhead}
\Cref{tab:runtime_token} reports the end-to-end overhead. In total, analyzing 20 apps takes \num{68.00} hours, \num{236,000} tool calls, and \num{120.80}M tokens. On average, this corresponds to \num{3.40} hours, \num{11,800} tool calls, and \num{6.04}M tokens per app. We also report time-to-first-vulnerability (TTV), which is \num{0.32} hours per app on average. At last, \system{} achieves an efficiency of \num{3.02} vulnerabilities per 1M tokens.

\input{token_cost}

\begin{tcolorbox}[
    colback=gray!10,     
    colframe=gray!50,    
    arc=0pt,             
    outer arc=0pt,
    boxrule=0.5pt,       
    left=5pt, right=5pt, top=2pt, bottom=2pt
]
\para{Takeaway for RQ1}
Multi-tool workflow vulnerabilities are common in today’s agent ecosystems: \system{} discovers \num{365} unique vulnerabilities across \num{19}/\num{20} agents, with \num{82.74}\% requiring multi-tool execution. Overall, \system{} achieves an efficiency of \num{3.02} vulnerabilities per 1M tokens.
\end{tcolorbox}

\subsection{RQ2: Individual Module Effectiveness}
\label{sec:rq2}

\input{precision_of_tool-chain}

\para{Tool-chain extraction accuracy}
We evaluate Module~A by measuring the extracted tool-chain graphs correspond to real inter-tool dependencies in the target agents. We use the same pool of candidate tool chains produced in RQ1 (\num{2,388} chains across 20 apps) and construct a validation set by uniformly sampling \num{10} chains per app (20$\times$10=\num{200} chains in total). For each sampled chain, we manually check every predicted edge using tool interfaces and code-level evidence of data flow (e.g., a returned field or persisted artifact that is subsequently consumed by the next tool). \Cref{tab:precision_chain_graph} reports two precision metrics: edge precision is \num{96.49}\% (\num{714}/\num{740} edges), and strict chain precision (a chain is correct only if \emph{all} edges are correct) is \num{91.50}\% (\num{183}/\num{200} chains). This indicates that the extracted chains used downstream are largely valid multi-tool workflows, rather than spurious tool combinations.

\input{effectiveness_of_TPS}

\para{TPS for reachability and stability}
We then evaluate TPS (Module~B) on a separate prompt-focused sample derived from the same extracted chains (200 chains in previous step). For each chain, we generate a seed prompt $P_0$ without TPS, and measure \emph{reachability} as the success rate of executing the target chain by the agent under $R{=}10$ repeated runs. We report reachability over $R{=}10$ independent runs to quantify nondeterminism, and separately report a stricter stability criterion (5/5 consecutive successes) to decide whether a prompt qualifies as a valid prompt. Next, we run TPS to obtain a \texttt{valid\_prompt} and re-measure reachability and stability, where stability is defined as success in 5/5 consecutive runs (the same criterion used to count \num{2,213} \texttt{valid\_prompt}s in Table~\ref{tab:overall_result}). Table~\ref{tab:tps_prompt_improvement} shows that TPS improves reachability from \num{27.05}\% to \num{95.45}\% (+\num{68.40}\%) and increases the stable prompt rate from \num{9.50}\% to \num{92.50}\% (+\num{83.00}\%). These gains indicate that prompt-level repair is necessary in practice even when the underlying tool chain is correct, because agents frequently fail due to missing preconditions, incorrect parameter bindings, or planner detours.

\input{effectiveness_of_payload_mutation}

\para{Guardrail-aware fuzzing}
Finally, we evaluate the guardrail-aware fuzzing component in Module~C using a vulnerability confirmed sample. We draw \num{20} verified vulnerabilities per sink type from the \num{365} unique vulnerabilities discovered in RQ1 (5 types $\times$ 20 = \num{100} vulnerabilities). For each vulnerability, we fix its corresponding \texttt{valid\_prompt} (5/5) and compare two vulnerability reproduction strategies: (i) \emph{initial payload} injection (without mutation), and (ii) \emph{mutation} using the strategies described in \Cref{subsubsec:Guardrail-aware fuzzing} (shard/encode/format). We report \emph{payload-level} trigger rates: each run with a concrete payload counts as one trial, and a trial is successful only when the sink-specific oracle is observed. For a matched budget, we perform five trials per malicious payload for both strategies. Concretely, the initial payload strategy repeats the same payload five times, while the mutation strategy uses five mutated payload variants (one run per variant). This yields \num{100} payload trials per sink type and \num{500} trials in total. 

\Cref{tab:trigger_rate_lift_by_sink_type} reports the resulting trigger rates, showing that mutation substantially improves reproduction, especially for guardrail-sensitive sinks such as CMDi and CODEi. As shown, initial payload injection triggers only \num{18.20}\% (\num{91}/\num{500}) of trials overall, whereas mutation achieves \num{88.60}\% (\num{443}/\num{500}), a \num{70.4}\% improvement. The lift is especially pronounced for guardrail-sensitive sinks such as CMDi (\num{10.00}\%$\rightarrow$\num{86.00}\%, +\num{76}\%) and CODEi (\num{7.00}\%$\rightarrow$\num{82.00}\%, +\num{75}\%), confirming that mutation is essential for reliable vulnerability reproduction under guardrails.

\begin{tcolorbox}[
    colback=gray!10,     
    colframe=gray!50,    
    arc=0pt,             
    outer arc=0pt,
    boxrule=0.5pt,       
    left=5pt, right=5pt, top=2pt, bottom=2pt
]
\para{Takeaway for RQ2}
The experiment result confirms that each module is effective in its intended role: Module~A extracts accurate tool chains, Module~B (TPS) turns them into valid and stable prompts, and Module~C (mutation) makes vulnerability reproduction reliable under guardrails.
\end{tcolorbox}

\subsection{RQ3: Ablation Study}
\label{sec:rq3_ablation}

\input{ablation_study}

We perform an ablation study to quantify how each component contributes to the final number of confirmed vulnerabilities, and to verify that the improvement of \system{} is not driven by a single step. \Cref{tab:ablation_chainfuzzer} reports the result in three column: (i) the number of unique vulnerabilities, (ii) the difference to the full system, and (iii) the number of sink types covered.

Overall, the largest drop comes from removing ChainExtraction: the results decrease from 365 to 63 ($-302$), and the covered sink types drop from 5 to 3.
This indicates that most findings require multi-tool composition rather than a single tool in isolation, because ChainExtraction is the step that exposes tool-to-tool dataflow edges needed to reach sink tools.
Removing TPS reduces the findings from 365 to 96 ($-269$) and reduces sink type coverage from 5 to 4, showing that many extracted chains are not directly reproducible without prompt stabilization; TPS increases the fraction of valid prompts that can consistently drive the agent along the intended chain. 
At last, removing Mutation yields 132 vulnerabilities ($-233$) while keeping all 5 sink types covered, which suggests Mutation mainly increases the number of confirmed instances within already reachable sink types rather than enabling new categories.

Taken together, the ablation results support the following interpretation: ChainExtraction primarily determines whether multi-tool sink reachability exists, TPS primarily determines whether the chain is executable and reproducible, and Mutation primarily determines how many vulnerabilities can be confirmed after reachability and validity are established.

\begin{tcolorbox}[
    colback=gray!10,     
    colframe=gray!50,    
    arc=0pt,             
    outer arc=0pt,
    boxrule=0.5pt,       
    left=5pt, right=5pt, top=2pt, bottom=2pt
]
\para{Takeaway for RQ3}
RQ3 confirms that the performance of \system{} is not attributable to any single component: removing ChainExtraction, TPS, or Mutation causes a substantial drop in confirmed vulnerabilities, showing the three modules are all necessary and complementary.
\end{tcolorbox}



\section{Discussion}
\label{sec:discussion}

\para{Limitations}
There are two limitations that affect the results in this paper. (1) Vulnerability reproducibility depends on the runtime environment (e.g., network egress policy, sandbox permissions, and availability of credentials), which can change whether certain chains are executable and whether a sink effect is observable; e.g., an SSRF chain may not trigger if outbound requests are blocked or if the target endpoint requires a token authentication that is not present. (2) \system{} uses an LLM in multiple modules (e.g., tool-chain extraction, TPS, and mutation), so errors from model hallucination or limited capability can prevent any module from reaching 100\% accuracy; although newer LLMs with stronger reasoning and lower hallucination rates are available, we use GPT-4o in our experiments to provide a conservative lower bound on performance. If a stronger model improves planning stability and reduces hallucinations, the end-to-end effectiveness of \system{} is expected to be detect more vulnerabilities.

\para{Mitigation strategy}
Our results suggest that defenses should be enforced at tool boundaries and sink invocations, because prompt wording alone is not a reliable control when multi-tool dataflow exists. In practice, developers can reduce risk with three engineering measures: (1) add sink-side input constraints so that dangerous tools do not accept free-form strings; e.g., for CMDi, replace \texttt{run(cmd)} with a fixed command template and validated arguments, and for SSRF, only allow requests to an allowlisted set of hostnames and URL schemes. (2) make cross-tool state \emph{explicit and mediated} rather than implicit and ambient, so that values can flow across tools only through typed artifacts with provenance; e.g., instead of letting a later tool read arbitrary workspace files or caches, require upstream tools to export outputs as structured objects (or signed/typed file handles) with metadata such as producer tool, schema type, and origin (user vs.\ external), and require downstream tools to consume only the declared fields/handles needed for the task. This preserves normal multi-tool workflows (e.g., retrieval $\rightarrow$ summarization $\rightarrow$ report generation) while preventing unintended data reuse (e.g., a web-fetched string being implicitly reused as a shell command). (3) implement lightweight dataflow-aware checks before calling sinks to block attacker-controlled content from reaching sink parameters; e.g., tag values derived from user input or untrusted external content and reject a call when such a value is used as a SQL query string in \texttt{execute} or as a template string in \texttt{from\_string}, forcing parameterization or escaping instead.

\section{Related Work}
\label{sec:related work}

\para{Vulnerabilities in LLM-based Systems}
The integration of Large Language Models (LLMs) into autonomous agents has expanded the attack surface for security vulnerabilities~\cite{related_work_LLM_1, related_work_LLM_2, related_work_LLM_3}. Early research focused on model-centric issues such as training data extraction and unintended capability misuse~\cite{related_work_LLM_4, related_work_LLM_5}, as well as inference-time attacks including membership inference and model inversion~\cite{related_inference_6_1,related_inference_6_2, related_inversion_7}. As agent frameworks began to call external tools, research shifted toward execution-time vulnerabilities~\cite{related_execution_time_8, related_indirect_injection_9, related_jailbreaks_10, related_malicious_tool_11, related_tool_poison_12}, including jailbreaks and prompt injection (both direct and indirect)~\cite{related_indirect_injection_9, related_jailbreaks_10}, malicious tool/plugin exploitation~\cite{related_malicious_tool_11, related_tool_poison_12}, and downstream effects such as command execution, SSRF, and unintended data access triggered through tool calls. In addition to integrity and control-flow risks, privacy leakage is also commonly studied in LLM-based systems, especially when agents have access to files, memory, or retrieval corpora. However, most existing works evaluate isolated attack scenarios or single-step behaviors, and do not provide a systematic workflow to discover and confirm vulnerabilities that are induced by multi-tool composition at scale. Recent work~\cite{zhao2025mindserversystematicstudy} has further shown that external MCP tool ecosystems can induce unintended privacy disclosure by composing content-ingestion, privacy-access, and outbound communication tools, whereas our work instead focuses on vulnerabilities inherently introduced by the agent's own multi-tool task-execution workflow.

\para{Vulnerability Discovery in Traditional Software}
Vulnerability discovery has a long history in software security, spanning Web applications, system software, mobile apps, and IoT platforms~\cite{related_work_web_1, related_work_web_2, related_work_web_3, related_work_mobile_1, related_work_mobile_2, related_work_mobile_3, related_work_iot_1, related_work_iot_2, related_work_iot_3}. A large body of work studies common vulnerability classes and their detection, such as injection vulnerabilities (e.g., command/SQL/template injection), SSRF, and insecure deserialization, using static and dynamic program analysis~\cite{related_work_program_analysis_1, related_work_program_analysis_2, related_work_program_analysis_3, related_work_program_analysis_4}. Fuzzing is another dominant approach, including coverage-guided fuzzing, grammar-based fuzzing, and hybrid fuzzing with symbolic or concolic execution~\cite{related_fuzzing_1,related_fuzzing_2,related_fuzzing_3,related_fuzzing_4}, where mutation and feedback signals are used to scale exploration and trigger deep bugs. Practical confirmation and triage often relies on oracles and reproducibility criteria (e.g., observable side effects, crash signatures, or controlled outputs), and many systems combine analysis with runtime monitoring such as taint tracking to validate that attacker-controlled data reaches sensitive sinks~\cite{related_binary_taint_1,related_dynamic_taint_1,related_dynamic_taint_2}. Our work draws on these classic ideas and adapts them to tool-augmented agent workflows, where vulnerabilities often emerge from composed tool interactions rather than a single API invocation.


\section{Conclusion}
\label{sec:conclusion}

This paper studies systemic workflow-level vulnerabilities in modern multi-tool LLM agents. We propose \system{}, a grey-box testing framework that uncovers and reproduces long-horizon vulnerabilities that emerge when multiple tools interact through intermediate artifacts (e.g., URLs, files, database records, and retrieved documents). In our evaluation on 20 popular open-source agent apps, \system{} discovers 365 unique vulnerabilities across 19 apps, where 82.74\% of the findings require multi-tool execution, highlighting that single-tool testing misses most workflow risks in practice. Overall, \system{} achieves an efficiency of 3.02 vulnerabilities per 1M tokens under practical end-to-end overhead, providing developers with actionable evidence to diagnose and mitigate compositional vulnerabilities in tool-augmented agent systems.


%% file: sink_type.tex
\begin{table}[htbp]
\caption{Candidate sink APIs used to label \texttt{sink\_tool} types in LLM Agent tools.}
\small
\centering
\resizebox{\linewidth}{!}{
\begin{tabular}{@{}l|c|l|c@{}}
\toprule
\textbf{Package} & \textbf{Class} & \textbf{Methods} & \textbf{Type} \\ \midrule
subprocess & / & \texttt{run}, \texttt{call}, \texttt{check\_call}, \texttt{Popen}, \texttt{getoutput} & CMDi \\
\cellcolor[HTML]{EFEFEF}os & \cellcolor[HTML]{EFEFEF}/ &
\cellcolor[HTML]{EFEFEF}\texttt{system}, \texttt{popen}, \texttt{exec*}, \texttt{spawn*} &
\cellcolor[HTML]{EFEFEF}CMDi \\
builtins & / & \texttt{eval}, \texttt{exec} & CODEi \\ \midrule
\cellcolor[HTML]{EFEFEF}urllib & \cellcolor[HTML]{EFEFEF}/ &
\cellcolor[HTML]{EFEFEF}\texttt{request.urlopen} &
\cellcolor[HTML]{EFEFEF}SSRF \\
requests & / & \texttt{get}, \texttt{post}, \texttt{request} & SSRF \\
\cellcolor[HTML]{EFEFEF}requests & \cellcolor[HTML]{EFEFEF}Session &
\cellcolor[HTML]{EFEFEF}\texttt{get}, \texttt{post}, \texttt{request} &
\cellcolor[HTML]{EFEFEF}SSRF \\
httpx & AsyncClient & \texttt{get}, \texttt{post}, \texttt{request} & SSRF \\
\cellcolor[HTML]{EFEFEF}aiohttp & \cellcolor[HTML]{EFEFEF}ClientSession &
\cellcolor[HTML]{EFEFEF}\texttt{get}, \texttt{post}, \texttt{request} &
\cellcolor[HTML]{EFEFEF}SSRF \\
urllib3 & PoolManager & \texttt{urlopen}, \texttt{request} & SSRF \\
\cellcolor[HTML]{EFEFEF}urllib3 & \cellcolor[HTML]{EFEFEF}/ &
\cellcolor[HTML]{EFEFEF}\texttt{request} &
\cellcolor[HTML]{EFEFEF}SSRF \\ \midrule
jinja2 & Environment & \texttt{from\_string} & SSTI \\
\cellcolor[HTML]{EFEFEF}flask & \cellcolor[HTML]{EFEFEF}Function &
\cellcolor[HTML]{EFEFEF}\texttt{render\_template\_string} &
\cellcolor[HTML]{EFEFEF}SSTI \\ \midrule
sqlite3 & Cursor & \texttt{execute} & SQLi \\
\cellcolor[HTML]{EFEFEF}sqlalchemy & \cellcolor[HTML]{EFEFEF}Session &
\cellcolor[HTML]{EFEFEF}\texttt{execute} &
\cellcolor[HTML]{EFEFEF}SQLi \\
sqlalchemy & Connection & \texttt{execute} & SQLi \\
\cellcolor[HTML]{EFEFEF}django & \cellcolor[HTML]{EFEFEF}/ &
\cellcolor[HTML]{EFEFEF}\texttt{cursor.execute} &
\cellcolor[HTML]{EFEFEF}SQLi \\ \bottomrule
\end{tabular}
}
\label{tab:sink_api_table_compact_lines}
\end{table}

%% file: Agent_app_experiment_object.tex
\begin{table}[htbp]
\caption{LLM agents in the dataset, ranked by GitHub stars.}
\small
\centering
\resizebox{\linewidth}{!}{
\begin{tabular}{@{}l|c|l|c@{}}
\toprule
\textbf{Agent} & \textbf{GitHub Stars} & \textbf{Agent} & \textbf{GitHub Stars} \\ \midrule
openclaw~\cite{openclaw} & 198k & superagi~\cite{superagi} & 17.2k \\
\cellcolor[HTML]{EFEFEF}autogpt~\cite{autogpt} & \cellcolor[HTML]{EFEFEF}182k & \cellcolor[HTML]{EFEFEF}agentscope~\cite{agentscope} & \cellcolor[HTML]{EFEFEF}16.4k \\
langflow~\cite{langflow} & 145k & agentzero~\cite{agentzero} & 14.7k \\
\cellcolor[HTML]{EFEFEF}langchain~\cite{langchain} & \cellcolor[HTML]{EFEFEF}127k & \cellcolor[HTML]{EFEFEF}bisheng~\cite{bisheng} & \cellcolor[HTML]{EFEFEF}11.1k \\
ragflow~\cite{ragflow} & 73.3k & xagent~\cite{xagent} & 8.5k \\
\cellcolor[HTML]{EFEFEF}metagpt~\cite{metagpt} & \cellcolor[HTML]{EFEFEF}64.2k & \cellcolor[HTML]{EFEFEF}taskweaver~\cite{taskweaver} & \cellcolor[HTML]{EFEFEF}6.1k \\
llama\_index~\cite{llama_index} & 47k & taskingai~\cite{taskingai} & 5.4k \\
\cellcolor[HTML]{EFEFEF}chatchat~\cite{chatchat} & \cellcolor[HTML]{EFEFEF}37.3k & \cellcolor[HTML]{EFEFEF}langroid~\cite{langroid} & \cellcolor[HTML]{EFEFEF}3.9k \\
vanna~\cite{vanna} & 22.7k & griptape~\cite{griptape} & 2.5k \\
\cellcolor[HTML]{EFEFEF}dbgpt~\cite{dbgpt} & \cellcolor[HTML]{EFEFEF}18.1k & \cellcolor[HTML]{EFEFEF}lagent~\cite{lagent} & \cellcolor[HTML]{EFEFEF}2.2k \\
\bottomrule
\end{tabular}
}
\label{tab:agent_github_stars}
\end{table}

%% file: Overall_result.tex
\begin{table*}[htbp]
\caption{Overall result of analyzed apps, tools, candidate chains, and discovered vulnerabilities.}
\small
\centering
\resizebox{0.85\linewidth}{!}{
\begin{tabular}{@{}c|c|c|c|c|c|ccc@{}}
\toprule
\textbf{\# Total App} &
\makecell{\textbf{\# App affected by}\\\textbf{vulnerability}} &
\textbf{\# Tools} &
\textbf{\# Sink tools} &
\makecell{\textbf{\# Candidate tool}\\\textbf{chains}} &
\makecell{\textbf{\# Valid}\\\textbf{prompt}} &
\multicolumn{3}{c}{\textbf{\# Vulnerability}} \\ \cmidrule(l){7-9}
& & & & & & \textbf{single-tool} & \textbf{multi-tool} & \textbf{Total} \\ \midrule
20 & 19 & 998 & 199 & 2388 & 2213 & 63 & 302 & 365 \\
\cellcolor[HTML]{EFEFEF}\makecell{\textbf{Average}\\\textbf{number per app}} &
\cellcolor[HTML]{EFEFEF}-- &
\cellcolor[HTML]{EFEFEF}49.9 &
\cellcolor[HTML]{EFEFEF}9.95 &
\cellcolor[HTML]{EFEFEF}119.4 &
\cellcolor[HTML]{EFEFEF}110.65 &
\cellcolor[HTML]{EFEFEF}3.15 &
\cellcolor[HTML]{EFEFEF}15.1 &
\cellcolor[HTML]{EFEFEF}18.25 \\
\bottomrule
\end{tabular}
}
\label{tab:overall_result}
\end{table*}

%% file: Overall_result_sink_type.tex
\begin{table}[htbp]
\caption{Breakdown of discovered vulnerabilities by injection source and sink type.}
\small
\centering
\resizebox{\linewidth}{!}{
\begin{tabular}{@{}l|c|c|c|c|c|c@{}}
\toprule
\textbf{Injection type} & \textbf{CMDi} & \textbf{CODEi} & \textbf{SSRF} & \textbf{SSTI} & \textbf{SQLi} & \textbf{Total} \\ \midrule
User-source & 82 & 53 & 44 & 17 & 29 & \textbf{\makecell{225\\(61.64\%)}} \\ \midrule
Env-source & 39 & 26 & 53 & 6 & 16 & \textbf{\makecell{140\\(38.36\%)}} \\ \midrule
\rowcolor[HTML]{EFEFEF}
\textbf{Total} &
\textbf{\makecell{121\\(33.15\%)}} &
\textbf{\makecell{79\\(21.26\%)}} &
\textbf{\makecell{97\\(26.57\%)}} &
\textbf{\makecell{23\\(6.30\%)}} &
\textbf{\makecell{45\\(12.33\%)}} &
\textbf{365} \\
\bottomrule
\end{tabular}
}
\label{tab:vuln_breakdown_source_type}
\end{table}

%% file: token_cost.tex
\begin{table}[htbp]
\caption{Runtime and token-cost statistics of \system{} over all dataset (20 agent apps).}
\small
\centering
\setlength{\tabcolsep}{7pt}
\renewcommand{\arraystretch}{1.18}
\resizebox{\linewidth}{!}{
\begin{tabular}{@{}l|c|c|c|c|c@{}}
\toprule
\textbf{Scope} &
\textbf{Wall time} &
\textbf{Tool calls} &
\textbf{Tokens} &
\textbf{TTV} &
\makecell{\textbf{Efficiency}\\\textbf{(vulns/1M tokens)}} \\ \midrule
Total & 68.00 h & 236{,}000 & 120.80 M & 6.40 h & \multirow{2}{*}{3.02} \\
\cmidrule(lr){1-5}
Average & 3.40 h & 11{,}800 & 6.04 M & 0.32 h & \\
\bottomrule
\end{tabular}
}
\label{tab:runtime_token}
\end{table}

%% file: precision_of_tool-chain.tex
\begin{table}[htbp]
\caption{Precision of extracted tool-chain graphs.}
\small
\centering
\resizebox{\linewidth}{!}{
\begin{tabular}{@{}c|c|c@{}}
\toprule
\textbf{\# Sampled chains} & \textbf{Edge Precision (\%)} & \textbf{Chain Precision (strict) (\%)} \\ \midrule
200 & \makecell{96.49\%\\(714/740 edges)} & \makecell{91.50\%\\(183/200 chains)} \\
\bottomrule
\end{tabular}
}
\label{tab:precision_chain_graph}
\end{table}

%% file: effectiveness_of_TPS.tex
\begin{table}[htbp]
\caption{Effectiveness of TPS on improving prompt validity metrics.}
\small
\centering
\resizebox{\linewidth}{!}{
\begin{tabular}{@{}l|c|c|c@{}}
\toprule
\textbf{Metric} & \textbf{Seed prompt} & \makecell{\textbf{Valid prompt}\\ \textbf{(After TPS)}} & \textbf{Improvement} \\ \midrule
Reach (R=10) & 27.05\% & 95.45\% & \textbf{+68.40\%} \\ \midrule
Stable prompt rate (5/5) & 9.50\% & 92.50\% & \textbf{+83.00\%} \\
\bottomrule
\end{tabular}
}
\label{tab:tps_prompt_improvement}
\end{table}

%% file: effectiveness_of_payload_mutation.tex
\begin{table}[htbp]
\caption{Trigger rate improvement from one-shot to mutation across sink tool types.}
\small
\centering
\resizebox{\linewidth}{!}{
\begin{tabular}{@{}l|c|c|c@{}}
\toprule
\makecell{\textbf{Sink tool} \\ \textbf{type}} & \makecell{\textbf{TriggerRate} \\ \textbf{(Initial payload)}} & \makecell{\textbf{TriggerRate} \\ \textbf{(Mutation)}} & \textbf{Improvement} \\ \midrule
\textbf{CMDi}  & \makecell{10.00\%\\(10/100)}  & \makecell{86.00\%\\(86/100)}  & 76\%   \\ \midrule
\textbf{CODEi} & \makecell{7.00\%\\(7/100)}    & \makecell{82.00\%\\(82/100)}  & 75\%   \\ \midrule
\textbf{SSRF}  & \makecell{29.00\%\\(29/100)}  & \makecell{91.00\%\\(91/100)}  & 62\%   \\ \midrule
\textbf{SSTI}  & \makecell{24.00\%\\(24/100)}  & \makecell{94.00\%\\(94/100)}  & 70\%   \\ \midrule
\textbf{SQLi}  & \makecell{21.00\%\\(21/100)}  & \makecell{90.00\%\\(90/100)}  & 69\%   \\ \midrule
\textbf{Total} &
\textbf{\makecell{18.20\%\\(91/500)}} &
\textbf{\makecell{88.60\%\\(443/500)}} &
\textbf{70.4\%} \\
\bottomrule
\end{tabular}
}
\label{tab:trigger_rate_lift_by_sink_type}
\end{table}

%% file: ablation_study.tex

\begin{table}[htbp]
\caption{Ablation study of \system{}: unique vulns and sink-type coverage.}
\small
\centering
\setlength{\tabcolsep}{7pt} 
\renewcommand{\arraystretch}{1.18} 
\resizebox{\linewidth}{!}{
\begin{tabular}{@{}l|c|c|c@{}}
\toprule
\textbf{Variant} & \textbf{Unique vulns} & \textbf{$\Delta$Vulns vs. Full} & \textbf{\# Sink types covered} \\ \midrule
\textbf{\system{}} & \textbf{365} & \textbf{0} & \textbf{5} \\ \midrule
w/o ChainExtraction & 63  & $-302$ & 3 \\ \midrule
w/o TPS & 96  & $-269$ & 4 \\ \midrule
w/o Mutation & 132 & $-233$ & 5 \\
\bottomrule
\end{tabular}
}
\label{tab:ablation_chainfuzzer}
\end{table}